\documentclass[twoside,a4paper,11pt]{article}
 
\usepackage{blindtext}

\usepackage{irrj2e}

\usepackage{lastpage}
\usepackage{amsmath}
\usepackage{color}
\usepackage{graphicx}
\usepackage{multirow}
\usepackage{subcaption}
\usepackage{url}
\usepackage{amsmath}
\usepackage{xcolor}
\usepackage{float}
\usepackage{afterpage}
\usepackage{comment}

\newcommand{\ContainedIn}{\ \mbox{$\vartriangleleft$}\ }
\newcommand{\Containing}{\ \mbox{$\vartriangleright$}\ }
\newcommand{\NotContainedIn}{\ \mbox{$\ntriangleleft$}\ }
\newcommand{\NotContaining}{\ \mbox{$\ntriangleright$}\ }
\newcommand{\BothOf}{\ \mbox{$\bigtriangleup$}\ }
\newcommand{\OneOf}{\ \mbox{$\bigtriangledown$}\ }
\newcommand{\Before}{\ \mbox{$\lozenge$}\ }

\newcommand{\GC}{\ \mbox{${\cal G}$\ }}

\newcommand{\Tau}{\mbox{$\tau$}\xspace}
\newcommand{\Rho}{\mbox{$\rho$}\xspace}

\newcommand{\Warren}{\texttt{Warren}\xspace}
\newcommand{\Tokenizer}{\texttt{Tokenizer}\xspace}
\newcommand{\Featurizer}{\texttt{Featurizer}\xspace}
\newcommand{\Annotator}{\texttt{Annotator}\xspace}
\newcommand{\Appender}{\texttt{Appender}\xspace}
\newcommand{\Idx}{\texttt{Idx}\xspace}
\newcommand{\Txt}{\texttt{Txt}\xspace}
\newcommand{\Hopper}{\texttt{Hopper}\xspace}

\irrjheading{1}{2025}{109-\pageref{LastPage}}{11/24; Revised 05/25}{05/25}{10.54195/irrj.19910}

\ShortHeadings{Annotative Indexing}{Clarke}
\firstpageno{109}

\begin{document}

\title{Annotative Indexing}

\author{\name Charles L. A. Clarke \email claclark@gmail.com \\
  \addr University of Waterloo\\
        Canada
}

\editor{Ismail Sengor Altingovde}

\maketitle

\begin{abstract}%
This paper introduces annotative indexing,
a novel framework that unifies and generalizes traditional inverted indexes,
column stores, object stores, and graph databases.
As a result, annotative indexing can provide the underlying indexing framework
for databases that support retrieval augmented generation, knowledge graphs, entity retrieval, semi-structured data, and ranked retrieval.
While we primarily focus on human language data in the form of text,
annotative indexing is sufficiently general to support a range of other data
types, and we provide examples of SQL-like queries over a JSON store that
includes numbers and dates.
Taking advantage of the flexibility of annotative indexing,
we also demonstrate a fully dynamic annotative index incorporating support for
ACID properties of transactions with hundreds of multiple concurrent readers
and writers.
\end{abstract}

\begin{keywords}
Search, Indexing, Inverted Indexes, Minimal-interval Semantics
\end{keywords}

\section{Introduction}

Until recently, and with few exceptions,
an inverted index provided the foundational file structure for
an information retrieval system.
Over the years,
research progress on file structures for information retrieval was primarily driven by the need to make traditional first-stage sparse retrieval methods (e.g., BM25) as fast as possible,
while minimizing storage and memory requirements,
motivating the development of specialized processing methods (e.g., WAND)
and compression methods (e.g., vByte).
To a large extent,
this research views an inverted index as single-purpose file structure,
with the sole task of delivering the top-$k$ items from a large collection to a second-stage re-ranker with high throughput and low latency.
More recently,
vector databases supporting dense retrieval have begun to replace inverted indexes,
but the focus remains on the efficiency and effectiveness of first-stage retrieval.

Managing large collections of human language data requires more than just a single-minded focus on first-stage retrieval.
For example, guidelines for the TREC 2024 RAG Track\footnote{
\url{https://trec-rag.github.io/}}
describe the preparation of a segmented version of the MS MACRO V2 passage corpus for use by track participants.
Processing steps include the identification and elimination of duplicate passages to avoid holes and inconsistencies in evaluation.
The original corpus was segmented with ``a sliding window size of 10 sentences and a stride of 5 sentences'' to make it ``more manageable for users and baselines.''
The original corpus and its de-duplicated/segmented version are distributed as two independent sets of compressed JSONL files,
linked to each other only by a naming convention for document identifiers.

In general, collections of human language data employ a variety of text formats, including JSON, JSONL, TSV, CSV, HTML, CBOR, LaTeX, Word, and PDF.
Even source code, such as Python and C++, can be considered as a form of human language data.
Processing text collections involves transformations such as tokenization,
sentence/word splitting, de-duplication, tagging, and entity linking,
as well as generating and storing weights for sparse retrieval and vectors for dense retrieval.
Tools for these tasks range from record-at-a-time processing in notebooks to storage in a variety of database systems,
including relational databases, search engines, object stores, and knowledge graphs.
No single tool allows us to flexibly store, transform, and search multi-format heterogeneous collections of unstructured and semi-structured human language data.

This paper introduces \textit{annotative indexing},
a novel framework that unifies and generalizes traditional inverted indexes,
column stores, object stores, and graph databases.
As a result, annotative indexing can provide the underlying indexing framework
for databases that support retrieval augmented generation, knowledge graphs, entity retrieval, semi-structured data, and ranked retrieval.
While we primarily focus on human language data in the form of text,
annotative indexing is sufficiently general to support a range of other data
types.
Annotative indexing facilitates dynamic update,
which in turn facilitates text processing pipelines that perform de-duplication, segmentation, and similar operations, expressing these operations by annotating the source text, rather than generating new text.

The next section (Section~\ref{sec:fun}) presents the fundamentals of annotative indexing, providing a foundation for the remainder of the paper.
Section~\ref{sec:related} places annotative indexing in the context of prior work.
Section~\ref{sec:org} then presents the overall organization of an annotative index.
As a proof of concept, the section also describes the architecture of our reference implementation, called Cottontail\footnote{
Code for the reference implementation is available at
\url{https://github.com/claclark/Cottontail}.
Following past practice in the information retrieval community,
the reference implementation is named after an animal,
in this case the eastern cottontail,
which is the most common species of rabbit in North America.
The author often encounters them out and about near the University of Waterloo.
}.
All experimental results in the paper were generated with this reference implementation.
The design of the reference implementation reflects the relative simplicity of
an annotative index, with a small number of generic components that can be
specialized and combined to support different applications.

Section~\ref{sec:processing} discusses query processing, including an example of a JSON store built on Cottontail, which supports structural containment, Boolean expressions, and similar operations, along with numbers and unified support for dates in differing formats.
Section~\ref{sec:update} discusses support for dynamic update and transactions, including support for ACID properties.
As an example,
Section~\ref{sec:update} presents a dynamically evolving collection that recapitulates the early years of TREC experiments,
with dozens of concurrent writers and hundreds of concurrent readers.
Annotative indexing and dynamic update complement and support each other.
Dynamic update of traditional inverted indexes is generally limited to adding and deleting entire documents, with no or limited support for concurrent update and transactions.
Annotative indexing provides the ability to annotate content after it has been added, enabling richer and more flexible update operations~---~~which, in turn, requires transactional support to ensure concurrency among multiple readers and writers.

\section{Fundamentals of Annotative Indexing}
\label{sec:fun}

\begin{figure}[t]
\vspace*{\baselineskip}
\includegraphics[width=\textwidth, keepaspectratio]{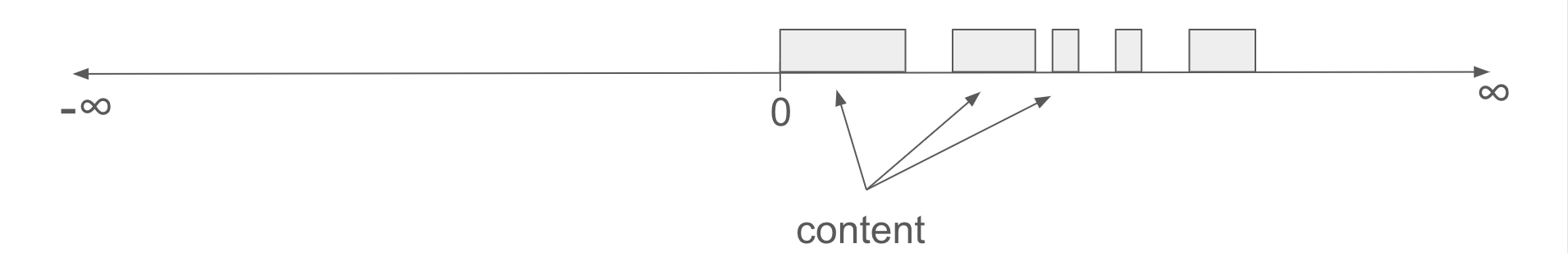}\\
\\
${\cal X}(592856130,592856138) \Rightarrow$
{\small\tt To be or not to be, | that is the}\\
\\
${\cal X}(17905274055,17905274393) \Rightarrow$
\vspace*{-0.5\baselineskip}
{\small
\begin{verbatim}
    { "docid": "msmarco_v2.1_doc_29_677149#3_1637632" , "end_char": 3061 ,
    "headings": "Aeolian Vibration of Transmission Conductors Aeolian
    Vibration of Transmission Conductors What is Aeolian Vibration?
    Wind causes a variety of motions on transmission line conductors.
    Important among them are How Aeolian Vibration Occurs?  Theory/Mechanism...
    ...that creates an alternating pressure imbalance causing the conductor to
    move up and down at a ninety-degree angle to the flow direction." ,
    "start_char": 1806 , "title": "Aeolian Vibration of Transmission Conductors",
    "url": "https://studyelectrical.com/2019/07/aeolian-vibration-..." }
\end{verbatim}
}

\caption{
The content of an associative index is situated in an address space,
which may contain gaps, where content has been deleted.
A translation function ${\cal X}(p, q)$ maps an interval in the address space
to its associated content.
The figure shows examples from an index containing the segmented version
of the MS MARCO V2.1 Document Corpus as used by the TREC 2023 RAG Track.
In this example, tokenization is word based.
At the content level, a JSON object is represented as a sequence of tokens, with special tokens representing JSON structural tokens ({\tt "}, {\tt :}, etc.).
Annotations on top of the content define structural elements.
For example, the annotation
$\langle \texttt{:title:}, (17905274368,17905274374) \rangle$
indicates the interval containing the title in this object.
}
\label{fig:content}
\vspace{0.2cm}
\end{figure}

An annotative index stores human language data as its \textit{content} 
plus a set of \textit{annotations} describing that content.
The content is represented by a sequence of \textit{tokens},
where each token is assigned an integer location in an
\textit{address space}, as illustrated in Figure~\ref{fig:content}.
If content has been deleted, gaps are possible.
By convention,
our reference implementation appends content at increasing addresses,
starting at zero.
However, negative addresses are supported for mathematical simplicity and consistency (see example below).
As shown in Figure~\ref{fig:content},
a translation function ${\cal X}(p, q)$ maps an interval in the address
space to the associated content.
${\cal X}(p, q)$ is undefined if $(p, q)$ contains a gap.
For content addressing purposes,
tokenization can be flexibly defined at the word or character level.
Separate and distinct tokenization and stemming can be also employed for specific
applications, e.g.\ ranking.

Annotations provide information about intervals over the content.
An annotation is a triple $\left<f, (p, q), v\right>$,
where $f$ is a \textit{feature},
$(p, q)$ is the interval over which the annotation applies,
and $v$ is the value of the feature over that interval,
which defaults to 0.
For convenience we define:
\begin{eqnarray}
\left<f, p, v\right>   & = & \left<f, (p, p), v\right>\\
\left<f, (p, q)\right> & = & \left<f, (p, q), 0\right>\\
\left<f, p\right> & = & \left<f, (p, p), 0\right>
\end{eqnarray}
For example, the annotation:
\[\langle \texttt{start\_char:}, (17905274359,17905274362), 1806\rangle\]
indicates that over the interval $(17905274359,17905274362)$
the feature \texttt{start\_char:} has the value 1806,
as shown in Figure~\ref{fig:content}.
The annotation:
\[\langle \texttt{tf:porter:aeolian}, 17905274055, 17 \rangle \]
indicates that the Porter-stemmed term ``\texttt{aeolian}'' appears
17~times in the JSON object starting at address 17905274055.
The annotation:
\[\langle \texttt{:}, (17905274055,17905274393)\rangle\]
indicates that the interval $(17905274055,17905274393)$ contains
a JSON object, as represented by the feature ``\texttt{:}''.
The annotation: 
\[\langle \texttt{aeolian}, 17905274369\rangle\]
indicates that the word ``\texttt{aeolian}'' appears at that address.
We can use annotations like these to implement BM25 ranking on a JSON store,
but the annotative index itself merely stores the content and its associated
annotations.
The interpretation of the annotations as term and document statistics is left to the ranking algorithm.

Annotations are indexed by feature,
with two access methods ($\tau$ and $\rho$) that both take an address $k$
in the address space and return the first annotation greater than or equal to
$k$, according to the start or end address of the interval.
\begin{eqnarray}
\label{eqn:tau}
f.\tau(k)\! & \!=\! & 
\!\left\{
  \langle (p, q), v \rangle,
    \ \mbox{where} \ \langle f, (p, q), v \rangle
    \text{\ is the annotation for $f$ with minimal $p \geq k$} 
\right\} \;\;\;\; \\
\label{eqn:rho}
f.\rho(k)\! & \!=\! &
\!\left\{
  \langle (p, q), v \rangle,
    \ \mbox{where} \ \langle f, (p, q), v \rangle
    \text{\ is the annotation for $f$ with minimal $q \geq k$} 
\right\} \;\;\;\;
\end{eqnarray}
To simplify index organization and facilitate index processing,
the set of annotations for a feature must follow
\textit{minimal-interval semantics} as defined in
prior work, including~\cite{bv16,bv18,cc00}, \cite{clarke96}, and \cite{ccb95a}.
Minimal interval semantics requires that no annotation for the same feature
can be contained in another,
but they can overlap.
If $\langle f, (p, q), v \rangle$ and $\langle f, (p', q'), v'\rangle$
are annotations for feature $f$,
then either $p < p'$ and $q < q'$,
or $p > p'$ and $q > q'$.
The annotations for $f$ are thus totally ordered~---~in the same order~---~by
their start and end addresses.
For mathematical simplicity and consistency,
we consider every feature $f$ to have the annotations $\langle f, (-\infty, -\infty), 0 \rangle$ and $\langle f, (\infty, \infty), 0 \rangle$.
\cite{bv18} describe a set of intervals under minimal-interval semantics as an
element of a ``Clarke–Cormack–Burkowski lattice''.
Clarke, Cormack, and Burkowski (1995) themselves call it a ``generalized concordance list''.
In this paper, the term ``annotation list'' implies an ordered set of annotations
under minimal-interval semantics.

\noindent
Prior work on minimal-interval semantics demonstrated their practical value as a method for expressing queries over heterogeneous collections of semi-structured data,
providing efficient support for containment,
boolean, merge, proximity, ordering, and other structural operators.
This paper extends this prior work in two ways,
which together substantially increase expressive power.
First, while prior work treated singleton intervals as the only atomic unit for
indexing purposes, we index intervals of any length.
For example, by indexing intervals we can run a sentence splitter over
the content and add annotations to the index indicating sentence boundaries.
Second, we associate a value with each interval,
which is preserved by containment and merge operations.
For example, we can compute terms statistics over the content and add
annotations to support ranked retrieval.
In this work, the atomic unit for indexing is an annotation,
comprising a feature, an interval and a value.
Operators combine annotations lists to produce annotation lists.

At its core, an annotative index is just a set of features with the $\tau$ and $\rho$ methods of Equations~\ref{eqn:tau} and~\ref{eqn:rho}, restricted by minimal interval semantics and as supported by the translation function ${\cal X}$.
The generality of annotative indexing lies in the flexibility of what the features represent, which can include term statistics, document structure, and links.
Since implementation details are hidden behind $\tau$, $\rho$, and ${\cal X}$, an annotative index can implemented with a wide variety of data structures.
For example, Cottontail provides both static and dynamic index formats that employ (nearly) distinct storage structures.

\cite{ccb95a} and ~\cite{clarke96} contain many examples illustrating queries and query processing.
Suppose we want to find all objects (``\texttt{:}'') with ``\texttt{aeolian}'' in their ``\texttt{:title:}''.
the $\tau$ and $\rho$ methods work together to provide an efficient solution.\\
\label{page:alg}
\begin{minipage}[t]{4in}
  \begin{tabbing}
  \ \ \ \ \ \ \ \ \ \ \ \=\ \ \ \=\ \ \ \= \ \ \ \= \ \ \ \= \\ \kill
    \> {\tiny 1}  \> $k \leftarrow 0$\\
    \> {\tiny 2}  \> ${\cal S} \leftarrow \{\ \}$\\
    \> {\tiny 3}  \> {\bf while} $k \ne \infty${\bf :}\\
    \> {\tiny 4}  \> \> $\langle (p, q), v \rangle \leftarrow \mbox{\texttt{:title:}}.\rho(k)$\\
    \> {\tiny 5}  \> \> $\langle (p', q'), v' \rangle \leftarrow \mbox{\texttt{aeolian}}.\tau(p)$\\
    \> {\tiny 6}  \> \> {\bf if} $q' \ne \infty$ {\bf and} $q' \leq q${\bf :}\\
    \> {\tiny 7}  \> \> \> ${\cal S} \leftarrow {\cal S} \cup \{\mbox{\texttt{:}}.\rho(q)\}$\\
    \> {\tiny 8}  \> \> \> $k \leftarrow q + 1$\\
    \> {\tiny 9}  \> \> {\bf else:}\\
    \> {\tiny 10} \> \> \> $k \leftarrow q'$\\
  \end{tabbing}
\end{minipage}\\
After execution, the set ${\cal S}$ contains a set of annotations corresponding to the required objects,
which is itself an annotation list.
Lines~4 and~5 generate the next candidate title and occurrence of ``aeolian''.
Line~6 determines if the candidate title contains  ``aeolian'',
and if so, we add the associated object to ${\cal S}$.
Line~10 is the key to efficiency, setting $k$ to skip titles that can't contain the next candidate ``aeolian''.
If ``aeolian'' is relatively rare, and clustered in a relatively small number of objects,
we may be able to avoid considering most titles.
For simplicity, this solution assumes that all titles are contained in an object, which may not be true if the context mixes different types of text.
As shown later, we would not normally materialize ${\cal S}$ as a set, but rather would provide ``lazy'' access through ${\cal S}.\tau$ and ${\cal S}.\rho$ implemented in terms of $\tau$ and $\rho$ for the ``\texttt{:}'', ``\texttt{aeolian}'', and ``\texttt{:title:}'' features.

\noindent
As another example, we can define access methods for a window of $n > 0$ tokens as
$\texttt{\#}n.\tau(k) \equiv \langle (k , k + n -1), 0 \rangle$ and
$\texttt{\#}n.\rho(k) \equiv \langle (k - n + 1  , k), 0 \rangle$.
In the example above, replacing ``\texttt{aeolian}'' with the feature \texttt{\#12} generates the set of objects with titles at least 12 tokens long.
While titles can no longer be skipped, windows are generated ``as needed'' to test against each title.
Note that $\texttt{\#12}.\rho(0) \equiv \langle (-11, 0), 0 \rangle$, providing a simple example of a negative address.
Section~\ref{sec:gcl} provides additional background on minimal interval semantics, including additional discussion regarding operators and query processing 

\section{Comparison with Prior Work}
\label{sec:related}

\subsection{Static and Dynamic Inverted Indexes}
\label{sec:inverted}

Annotative indexes generalize inverted indexes.
\cite{bcc10} provides a review of inverted index file structures and
associated query processing methods that remains reasonably current.
They describe the core techniques that are still widely employed,
along with experimental comparisons against competing techniques.
A generic inverted index maps each term in a 
\textit{vocabulary}~---~maintained in by a \textit{dictionary}~---~to a
\textit{postings list} of \textit{document identifiers} where the term appears.
Postings lists often include term frequencies to support ranking formulae
and term offsets to support phrase searching.
Postings lists are typically gap-encoded and compressed with a method such as 
vByte \citep{vbyte}, which usually provides an acceptable trade off between
compression ratio and decompression speed.
Since query processing methods can often skip documents, synchronization points
may be included in the compressed posting lists to improve performance
and reduce the need for decompression \citep{mz96}.

Academic research on inverted indexes often views them through the lens of a static file structure, built once from a collection and never changed (e.g., \cite{arroyuelo2018hybrid,10.1007/978-3-030-15712-8_23,10.1145/3340531.3412000}).
If the collection changes, the index is re-built from scratch.
For example, if a researcher wants to remove near-duplicates from a collection because they are causing problems with their retrieval experiments, the researcher first filters the collection and then builds a new index for the filtered collection.
A complete index re-build can be slow, even for a relatively small collection.
At the very least, a complete re-build requires an end-to-end read of the
collection to construct postings lists,
so that the build time grows linearly with the size of the collection.

Prior research has considered a variety of dynamic update models for inverted indexes \citep{bcc10}.
The simplest model provides for \textit{batch updates},
which build index structures for new documents and merge them into the original index without requiring a complete rebuild.
During the merge, the index also deletes any unneeded documents.
Ideally, the overall process is managed as a transaction,
so that a failure during the update process does not corrupt the index structures.
The batch update model supports only one transaction at a time.
Starting a second update during a transaction either produces an error or blocks until the current transaction completes.
If the index is queried during a transaction,
atomicity should guarantee that a result over the original index is returned.
While a batch update avoids some of the work required by a full rebuild,
the index cannot evolve quickly.
Depending on the final size of the index, it might take minutes or hours for a change to become visible to queries.

\noindent
Under the  \textit{immediate-access} dynamic update model, 
changes become visible as soon as they are made \citep{bcc10}.
Social media search provides an important use case for immediate-access dynamic update.
\cite{10.1145/2487575.2488221} describe index update in the EarlyBird search engine,
developed for Twitter.
EarlyBird search was designed for a single, high-volume update stream, with many concurrent readers and a strong temporal ranking signal, placing a high priority on recent tweets.
Once a tweet is indexed, its indexing does not change.
\cite{MOFFAT2023103248} explore trade-offs between insertion speed, query speed, and index size in an immediate-access dynamic index. They describe in-memory indexing structures that supports a stream of interleaved queries and document insertions.
Document deletion is not supported, so that index grows with each insertion.
To maintain a consistent view of the index, they assume that ``all postings associated with each ingested document are processed into the index before the next query operation is permitted'', effectively requiring a read-lock on the index during each insertion.
\cite{10.1007/978-3-030-99736-6_11} describe index structures for dynamic update that uses fixed volume of memory, with older documents expiring from the index.

Prior research on immediate-access dynamic update does not satisfy the requirements of annotative indexing.
In particular, prior work assumes that indexing for a document happens all at once;
no additional indexing for a document can be added at a later time \citep{bcc10}.
In contrast, annotative indexing enables novel use cases that require additional indexing.
For example, imagine an annotative index supporting a document ingestion pipeline for a retrieval augmented generation (RAG) system that includes de-duplication, segmentation, and indexing stages.
Each stage reads its input from the index and records its output by adding annotations.
For an annotative index to fully support a document processing pipeline, the output from a
stage must be immediately visible as soon as the stage finishes.
However, each stage must see a complete and consistent view of the output from the previous stages.
When stages are independent of each other, it should be possible for them to run concurrently.
Since some stages may require considerable processing, it is important for updates to be durable, allowing the pipeline to recover quickly from a failure.
To satisfy this scenario and realize the full benefits of annotative indexing, an annotative index must support concurrent access and ensure ACID properties of transactions, requirements that are not met by prior research.

\subsection{First-Stage Retrieval}
\label{sec:first}

Over 30 years after its invention, the BM25 formula remains the touchstone
for unsupervised first-stage retrieval \citep{rw94,rwjhg94}.
When compared to other unsupervised retrieval formulae from the 1990s and early
2000s~---~which may provide as-good-or-better retrieval effectiveness~---~BM25
exhibits term saturation properties that can be exploited to
substantially improve query performance through WAND query processing
\citep{wand,magic,tf95} and Block-Max WAND processing  \citep{dns13,ding2011faster}.
Term saturation guarantees an upper bound on the weight given to any single
query term, allowing us to skip documents whose score cannot exceed a
threshold defined by scores of the current top-$k$ documents.

\noindent
In their description of standard WAND processing, \cite{magic} assume
posting lists will be accessed through three functions:
1) A \textit{first} function, which creates an iterator for the list,
2) a \textit{next} function,
which advances the iterator by one posting, and
3) a \textit{seek(d)} function, which advances iterator to the first document identifier greater than or equal to \textit{d}.
The \Tau and \Rho operations in Equation~\ref{eqn:tau} and
Equation~\ref{eqn:rho} generalize the \textit{seek} function to intervals.
When wrapped in an appropriate iterator, and with appropriate annotations,
they directly support WAND processing over annotative indexes.

Using the \Tau and \Rho operations, Cottontail can efficiently implement WAND processing.
As a demonstration of performance we use the 6980 ``dev small’’ queries and passages from the original MS MARCO test collection \citep{marco}.
The corpus comprises 8,841,823 passages with an uncompressed size of 2.85GB. Full cottontail indexing with BM25 annotations using the static index implementation described in Section~\ref{sec:org} gives a compressed index of 4.4GB.
In addition to the annotations required for BM25 ranking, this index includes token-level annotations, to implement phrase search and structural queries, along with the full text of the collection, to implement the ${\cal X}$ translation function.

With BM25 parameters $b = 0.68$ and $k_1 = 0.82$ we achieve an MRR@10 of 0.185.
These BM25 parameters are recommended by the Anserini onboarding guide, which uses this collection as an example for teaching indexing and retrieval\footnote{
\url{ https://github.com/castorini/anserini/blob/master/docs/experiments-msmarco-passage.md}}.
MMR@10 is the standard precision metric for this test collection.
Running with two threads per physical core, i.e., 46 concurrent queries, it requires less than 20 seconds to rank all queries to a depth of 10, giving a throughput of over 350 queries/second\footnote{
All experiments reported in this paper were conducted on a
Intel(R) Xeon(R) Gold 5120 CPU with 256GB of memory.
}.
Running one query at a time gives an average query latency of 65ms. In comparison, using a system based on the Lucene search library, \cite{lyl20} report a BM25 latency of 55ms and MMR@10 of 0.184 on the same collection. While \cite{lyl20} do not indicate the hardware used for their measurements, nor do they report query throughput, this comparison suggests that our general indexing framework can be reasonably competitive with a specialized index developed through years of engineering effort.

In recent years, neural retrieval methods have eclipsed traditional unsupervised methods for first-stage retrieval.
Neural first-stage retrieval methods fall into two camps: \textit{sparse vector retrieval} and \textit{dense vector retrieval}.
Sparse vector retrieval methods represent queries and documents in a high-dimensional space, where each dimension corresponds to a token \citep{uniCOIL, WordPiece} and most weights are zero, especially in query vectors.
Sparsity allows these vectors to be stored in an inverted index; ranking requires only a dot product between the query and document vectors.
Successful approaches to sparse retrieval include DeepCT \citep{DeepCT}, HDCT \citep{HDCT}, uniCOIL \citep{uniCOIL} and SPLADE \citep{splade}.
In particular, SPLADE is widely recognized for its retrieval effectiveness \citep{splade2,BlockMax24,EffSparse24}.
Despite a few proposals for unsupervised neural sparse methods (e.g., \cite{bm26}),
neural sparse methods are often called ``learned sparse retrieval'' methods to distinguish them from traditional unsupervised sparse methods, such as BM25.
 
Annotative indexing trivially supports learned sparse retrieval by creating an annotation for each element of a sparse vector.
It is also trivial to support multiple sparse retrieval methods (e.g. BM25 and SPLADE) in the same index, or to use different ranking approaches at different structural levels (e.g. BM25 at the document level and SPLADE at the passage level) .
Unfortunately, learned weights do not provide the distributional properties that algorithms like WAND exploit to improve query performance.
Score-at-a-time ranking approaches can partly address this problem \citep{wacky}.
It may also be possible to adapt block pruning and other methods to annotative indexes, for example, by adding additional annotations summarizing weights over blocks of documents \citep{BlockMax24,EffSparse24,10.1145/3077136.3080780,ding2011faster}.

Dense vector retrieval is currently the focus of intense research, with multiple recent surveys available \citep{denseS0,denseS1}.
The simplest form of dense retrieval, \textit{bi-encoder retrieval}, represents queries and documents in a low-dimensional space (e.g., 768 dimensions) where the values in most dimensions are non-zero. Ranking requires only a dot product between the query and document vectors \citep{denseA,denseB,denseC}.
Various approximate k-nearest neighbor search methods can speed the ranking process. For example, Hierarchical Navigable Small World (HNSW) graphs arrange vectors in a hierarchy of proximity graphs that can be traversed in approximately logarithmic time \citep{hnsw}.

While each dimension could be represented as an annotation list, retrieval would be inefficient because of the relatively large number of non-zero values in a dense vector.
To support dense vectors, we would need to extend annotative indexing with a vector store, which might map locations in the address space to vectors.
Annotative indexes can also help support hybrid approaches that combine sparse and dense retrieval \citep{forward}.
It may also be possible to encode and efficiently search HNSW graphs encoded as annotations.

\subsection{Minimal Interval Semantics}
\label{sec:gcl}

Minimal-interval semantics were invented by the author for his Ph.D
thesis nearly 30 years ago \citep{clarke96}.
If we view the result of a text search over a string as a set of substrings
that satisfy the requirements of the search,
minimal-interval semantics provide a simple and natural way to linearize the
set, as well as enabling fast and flexible algorithms for combining and
filtering search results.
If we specify the set of substrings $S$ as a set of intervals $(p, q)$,
minimal interval semantics allows these intervals to overlap but not to nest.

An interval~\mbox{$(p,q)$} {\em overlaps\/} an interval~\mbox{$(p',q')$}
if either \mbox{$p' \leq p \leq q'$} or \mbox{$p' \leq q \leq q'$},
but not both.
An interval~\mbox{$(p,q)$} is {\em nested\/} in an interval~\mbox{$(p',q')$} if
\mbox{$(p, q) \neq (p', q')$} and \mbox{$p' \leq p \leq q \leq q'$}.
If~\mbox{$a = (p,q)$} and~\mbox{$b = (p',q')$} are intervals, the
notation~\mbox{$a \sqsubset b$} indicates that $a$ nests
in $b$; the notation~\mbox{$a \sqsubseteq b$} indicates that $a$ is
{\em contained in\/} $b$: that either $a$ and $b$ are equal or that $a$
nests in $b$.
Intervals form a partial order under $\sqsubseteq$.

We formalize the reduction of a set of intervals $S$ to a
\textit{generalized concordance list} as a function~$\mbox{\GC(S)}$:
\[
  \GC(S) =
    \{
      a\ |\ a \in S\ {\rm and} \not\!\exists\ b \in S\
      {\rm such\ that}\ b \sqsubset a
    \}
\]
A set $S$ is a generalized concordance list if and only if
$S =\!\GC (S)$.
Each interval in a generalized concordance list acts as a ``witness'' to the
satisfiability of the requirements of the search \citep{bv18}.
As a simple example, consider the query:
\begin{center}
\texttt{"peanut butter"} \BothOf \texttt{"jelly doughnut"},
\end{center}
where ``\BothOf'' indicates Boolean conjunction.
If we view the set of intervals satisfying the query \texttt{"peanut butter"}
as the set of all intervals containing that string, of any length,
then $\GC(\texttt{"peanut butter"})$
is just the set of intervals corresponding to the string itself.
If we view the set of intervals satisfying the conjunction as the set of
intervals that contain both strings, 
then the set of minimal intervals that contain both strings is
$\GC(\texttt{"peanut butter"} \BothOf \texttt{"jelly doughnut"})$,
which may overlap but not nest.
For example, the sentence:
\begin{center}
\textit{
Peanut butter on a jelly doughnut is better than a peanut butter sandwich.
}
\end{center}
contains two overlapping intervals which satisfy the conjunction under
minimal interval semantics.

\begin{figure}[t]

\begin{tabbing}
\ \ \ \ \= \ \ \ \ \=\kill

{\bf Containment Operators}\\

\> Contained In:\\
\> \> $
  A \ContainedIn B =
      \{
        a\ |\ a \in A\ {\rm and}\ \exists\ b \in B\ {\rm such\ that}\
        a \sqsubseteq b
      \}
$\\

\> Containing:\\
\> \> $
  A \Containing B =
      \{
        a\ |\ a \in A\ {\rm and}\ \exists\ b \in B\ {\rm such\ that}\
        b \sqsubseteq a
      \}
$\\

\> Not Contained In:\\
\> \> $
  A \NotContainedIn B =
      \{
        a\ |\ a \in A\ {\rm and} \not\!\exists\ b \in B\
        {\rm such\ that}\ a \sqsubseteq b
      \}
$\\

\> Not Containing:\\
\> \> $
  A \NotContaining B =
      \{
        a\ |\ a \in A\ {\rm and} \not\!\exists\ b \in B\
        {\rm such\ that}\ b \sqsubseteq a
      \}
$\\

\\

{\bf Combination Operators}\\
\> Both Of:\\
\> \> $
  A \BothOf B =
    \GC(
      \{
        c\ |\
          \exists\ a \in A \ {\rm such\ that}\ a \sqsubseteq c
          \ {\rm and}\ \exists\ b \in B \ {\rm such\ that}\ b \sqsubseteq c
      \}
    )
$\\

\> One Of:\\
\> \> $
  A \OneOf B =
    \GC(
      \{
        c\ |\
          \exists\ a \in A \ {\rm such\ that}\ a \sqsubseteq c
          \ {\rm or }\ \exists\ b \in B \ {\rm such\ that}\ b \sqsubseteq c
      \}
    )
$\\

\> Follows:\\
\> \> $
  A \Before B =
    \GC(
      \{
        c\ |\
          \ \exists\ (p,q) \in A \ {\rm and}
          \ \exists\ (p',q') \in B \ {\rm where}
          \ q < p'\ {\rm and}\ (p, q') \sqsubseteq c
      \}
    )
$
\end{tabbing}

\caption{
Fundamental operators for expressing structural relationships over generalized
concordance lists, which underlie annotation lists.
$A$ and $B$ can be any generalized concordance lists, including subqueries built from these operators.
}
\label{fig:operators}
\vspace{0.4cm}
\end{figure}

Figure~\ref{fig:operators} summarizes fundamental operators from
\cite{clarke96}, where $A$ and $B$ are generalized concordance lists,
The operators fall into two groups,
\textit{containment} and \textit{combination},
which together support a wide range of queries specifying structural
relationships.
For example, the query for all objects (``\texttt{:}'') with ``\texttt{aeolian}'' in their ``\texttt{:title:}'' is:
\begin{equation*}
\texttt{:} \Containing (\texttt{:title:} \Containing \texttt{aeolian})
\end{equation*}
Generalized concordance lists have the same access methods as annotation lists,
but without associated values.
They are just ordered sets of intervals under mimimal interval semantics.
If $S$ is a generalized concordance list, then:
\begin{eqnarray}
\label{eqn:gcltau}
S.\tau(k) & = &
\left\{
  (p, q), \ \mbox{where} \ (p, q)\ \text{\ is the interval in $S$ with minimal $p \geq k$}
\right\}\\
\label{eqn:gclrho}
S.\rho(k) & = &
\left\{
  (p, q), \ \mbox{where} \ (p, q)\ \text{\ is the interval in $S$ with minimal $q \geq k$}
\right\}
\end{eqnarray}
A key observation of \cite{clarke96}~---~echoed by \cite{bv16}~---~is
that evaluation can be ``lazy''.
Much like WAND processing, lazy evaluation allows us to skip
solutions to subqueries that cannot lead to a solution for the overall
query.
If $A$ and $B$ are generalized concordance lists,
we can implement \Tau and \Rho access methods for each operator of Figure~\ref{fig:operators} in terms of \Tau and \Rho for $A$ and $B$,
which in turn can subqueries built from these operators.
For example,
the \Rho operator for $A \Containing B$ can be written as:\\
\begin{minipage}[t]{4in}
  \begin{tabbing}
    \ \ \ \ \ \ \ \ \ \ \ \=\ \ \ \=\ \ \ \= \ \ \ \= \ \ \ \= \\ \kill
    \> {\tiny 1}  \> $(A \Containing B).\rho(k) \equiv$ \\
    \> {\tiny 2}  \> \> $(p , q ) \leftarrow A.\rho(k)$ \\
    \> {\tiny 3}  \> \> $(p', q') \leftarrow B.\tau(p)$ \\
    \> {\tiny 4}  \> \> {\bf if} $q' \leq q${\bf:} \\
    \> {\tiny 5}  \> \> \> {\bf return} $(p,q)$ \\
    \> {\tiny 6}  \> \> {\bf else:} \\
    \> {\tiny 7}  \> \> \> {\bf return} $(A \Containing B).\rho(q')$\\
  \end{tabbing}
\end{minipage}

If $A$ corresponds to ``\texttt{:title:}'' and $B$ corresponds to ``\texttt{aeolian}'', we obtain a \Rho access method for titles containing ``aeolian'', which can in turn be combined with the generalized concordance list for objects to give a \Rho access method for objects containing these titles.
Compare this definition to the algorithm on page~\pageref{page:alg}.

Unfortunately, there is no single summary of key theorems and algorithms for
operations under minimal interval semantics.
\cite{ccb95b} contains an overview with many examples.
\cite{ccb95a} contains additional examples and algorithms for implementing the
operators, which are extended with proofs by \cite{clarke96}.
While \Tau and \Rho are sufficient to implement the operators of
Figure~\ref{fig:operators}, \cite{clarke96} also defines ``backwards'' version
of these access methods that facilitate solutions that start with the last
interval,
which can be valuable in finding the most-recent solutions to queries over
a growing index \citep{10.1145/2487575.2488221}.
More recently, \cite{bv16,bv18}  present a mathematical foundation
for minimal-interval semantics based on lattices.

\cite{cc00} present an improved implementation framework for the combinational 
operators, including Boolean operators.
Under this framework,
finding all solutions to a combinational query with $n$ terms requires
no more than $O(n \cdot {\cal A})$ calls to access methods for the terms,
where ${\cal A}$ is the number of \textbf{solutions} to the query.
Using \textit{galloping search} \citep[pp.\ 42--44]{bcc10}
to implement the access methods for the terms gives an overall time
complexity of $O(n \cdot {\cal A} \cdot \log{(L/{\cal A})})$,
where $L$ is the length of the longest posting list for a term.
The overall time complexity is nearly linear in the number of solutions,
rather than the length of any postings list, as might be expected.
If there are few solutions, most of the postings lists might be skipped.
The reference implementation for annotative indexing, Cottontail,
captures most of the cumulative insights from research on minimal
interval semantics\footnote{
\url{https://github.com/claclark/Cottontail/blob/main/src/gcl.cc}
}.

\subsection{Column Stores}

A column store is a physical design strategy for relational database systems
that partitions data primarily by column, rather than by row. 
As opposed to a traditional row-oriented strategy for physical database design,
a column-oriented strategy is known to provide better performance on
data analytics and other read-intensive workloads.
Among other properties, grouping together values from a single column can
improve compression because values within the same column are often
similar or repetitive.
This homogeneity makes it easier to apply compression techniques that exploit
redundancy.
The popularity of column stores grew from the success of systems such as
C-Store \citep{cstore} and MonetDB \citep{monet}.
Currently, open formats such as ORC and Parquet enable support for
columnar storage on most platforms for data analytics.

Inverted indexes are close cousins of column stores \citep{olddog}.
If we consider the terms in the vocabulary as columns of a table whose rows
represent documents,
then we can imagine the table as containing term weights,
perhaps with a special column containing document identifiers.
Since most terms appear only in a relatively small number of documents,
the table is sparse.
Most of the entries are NULL.
Computing a retrieval formula requires an aggregation over the columns
corresponding to terms in the query.
Since an inverted index organizes this table by column,
only query columns need to be accessed to compute the retrieval formula.

Annotative indexes generalize inverted indexes to something close to
a column store.
If we store rows of a table as the content of the annotative index and 
treat the features as columns,
then each annotation represents an entry in the table:
\[
  \langle
    \textit{column}, (\textit{start}, \textit{end}), \textit{value}
  \rangle,
\]
where $(\textit{start}, \textit{end})$ is the location of the value in the
row.
Data can be accessed by column through annotation lists,
and by row through the translation function ${\cal X}(p, q)$.

\subsection{Graph Structures}
\label{sec:links}

Many application areas now require database support to store and process large graphs structures \citep{sahu2023ubiquity}.
For example, Facebook developed the Unicorn search engine to store and search social graph information at worldwide scale \citep{unicorn}.
Unicorn's core file structures extend inverted lists with additional information, similar to annotative indexing.
However, it does not support minimal-interval semantics.
In an information retrieval context, graph data often takes the form of a \textit{knowledge graph}, with \cite{bast2025knowledge} providing a current survey.
Knowledge graphs often encode relationships as subject-predicate-object triples:
For example the triple:
\[
  \langle \texttt{Meryl\_Streep} \rangle-\langle \texttt{won\_award} \rangle-\langle \texttt{Best\_Actress}\rangle
\]
indicates that Meryl Streep won an Oscar for Best Actress.

An annotation list can encode a directed graph by storing a location in the address space as the value of annotation, so that the annotation $\left<G, p, v\right>$ is interpreted as a link from an object containing the location $p$ to an object containing the location $v$.
For example, consider the trivial friend graph:
\begin{verbatim}
               {"name": "Alice", "friends": ["Bob", "Carol", "Dave"]}
               {"name": "Bob", "friends": ["Alice", "Dave"]}
               {"name": "Carol", "friends": ["Alice"]}
               {"name": "Dave", "friends": ["Bob", "Alice"]}
\end{verbatim}
If the Alice object is stored at $(0,26)$ and the Bob object is stored at $(27,49)$,
the annotation $\langle \texttt{@friend}, 7, 27 \rangle$ indicates a link from Alice's friends array to Bob, where $7$ is the address of the token ``Bob'' the occurs in Alice's friend array. 
Using a similar approach, annotations can encode subject-predicate-object triples.
\[
  \langle
    \textit{predicate}, \textit{subject}, \textit{object}
  \rangle.
\]
To encode a triple indicating that Meryl Streep won an award for best actress,
the \textit{predicate} would be encoded as the feature \texttt{won\_award},
the \textit{subject} would be an address in the record associated with Streep,
and the \textit{object} would be an address in the record associated with the best actress award.

\section{Organization of an Annotative Index}
\label{sec:org}

In this section, we consider the organization and construction of an
annotative index, using our reference implementation, Cottontail, as an example.
Cottontail provides two distinct implementations of the index structures,
a \textit{static index} and a fully \textit{dynamic index}.
The static index supports larger collections,
where it may not be possible to maintain the entire collection in memory.
The  static index reads annotation lists from storage only for query processing and index update;
it supports only a single update transaction at a time under the batch update model.
The dynamic index  maintains all active index structures in memory,
while still durably committing transactions to storage.
In this section, we focus on the basic index construction process,
which applies to both static and dynamic indexes.
Section~\ref{sec:update} extends this material with details for the fully dynamic index,
including support for immediate update and multiple concurrent readers and writers.

An annotative index extends and generalizes an inverted index,
as outlined in Section~\ref{sec:inverted},
annotations are indexed by feature, with annotations ordered by the start
address (and equivalently the end address) of their intervals.
If the annotations for feature $f$ are
\[
a_0 = \left<f, (p_0, q_0), v_0\right>,\ 
a_1 = \left<f, (p_1, q_1), v_1\right>,\ 
a_2 = \left<f, (p_2, q_2), v_2\right>,...
\]
then $\forall i,\ p_i < p_{i+1}$ and $q_i < q_{i+1}$.
Since they strictly increase,
successive start (and end) addresses can be gap-encoded
and compressed with vByte,
or other methods developed for compressing postings lists.
For a given $f$, if $\forall i,\ p_i = q_i$,
then its end addresses can be compressed away.
Similar to column stores, values will tend to share distributional 
properties that can be exploited to improve compression.
For a given $f$, if $\forall i,\ v_i = 0$, its values can be compressed away.

Figure~\ref{fig:cottontail} provides an overview of the major components
of Cottontail.
The various components of an annotative index are grouped into a \Warren,
which manages transactions and simplifies common operations that interact
with multiple components\footnote{
The author is aware that eastern cottontail rabbits are solitary and don't
live in warrens.
}.
Apart from a \Warren,
each component implements no more than three operations.
Cottontail provides multiple versions of each component,
each specialized for a different purpose,
which can be mixed and matched in a \Warren.

\begin{figure}[t]

\begin{tabbing}
\ \ \ \ \= \ \ \ \ \=\kill
\texttt{Warren:} Groups the following components and manages transactions.\\
\\
\>Operations: 
\texttt{clone},
\texttt{start},
\texttt{end},
\texttt{transaction},
\texttt{ready},
\texttt{commit},
\texttt{abort}\\
\\
\>\texttt{Tokenizer:}Facilitates content addressability (Section \ref{sec:org}).\\
\>\>Operations: \texttt{tokenize}, \texttt{split}, \texttt{skip}\\
\\
\>\texttt{Featurizer:}
  Maps a feature (expressed as a string) to a 64-bit value (Section \ref{sec:org}).\\
\>\>Operations: \texttt{featurize}\\
\\
\>\texttt{Annotator:} Inserts and deletes annotations (Section \ref{sec:org}).\\
\>\>Operations: \texttt{annotate}, \texttt{erase}\\
\\
\>\texttt{Appender:} Appends text to the content (Section \ref{sec:org}).\\
\>\>Operation: \texttt{append}\\
\\
\>\texttt{Idx:} Provides read access to annotations (Section \ref{sec:processing}).\\
\>\>Operation:
  \texttt{hopper(f)}~---~create
  a cursor (called a \texttt{Hopper}) for the feature \texttt{f}\\
\\
\>\texttt{Txt:} Provides read access to content (Section \ref{sec:processing}).\\
\>\>Operation:
  \texttt{translate(p, q)}~---~return
  content associated with the interval (${\cal T}(p, q)$)\\
\end{tabbing}
\caption{
Major components of Cottontail,
the reference implementation for annotative indexing.
A \texttt{Warren} object contains and manages one instance of each of the other components
(\texttt{Tokenizer}, \texttt{Featurizer}, etc.). 
Cottontail provides multiple versions of each component,
each specialized for a different purpose.
Section numbers indicate where the component is discussed.
The transaction model for a \texttt{Warren} is discussed in Section~\ref{sec:update}.
}
\label{fig:cottontail}
\vspace{0.4cm}
\end{figure}

A \Tokenizer facilitates content addressability by
splitting strings into tokens,
computing token boundaries,
and skipping tokens.
Support for ASCII content with HTML-style tags is provided by
\texttt{AsciiTokenizer},
which is intended for use with older TREC collections.
Generic support for Unicode is provided by \texttt{Utf8Tokenizer},
which is intended for use with JSON and other modern content.
The role of a \Tokenizer in a \Warren is limited to facilitating content
addressability.
Other tokenization (e.g., language specific or WordPiece) can be used by
features in annotations to support ranking and other applications.

Internally, cottontail represents an annotation as four 64-bit values,
using a \Featurizer to map a feature expressed as a string to a 64-bit value.
\texttt{HashingFeaturizer} maps strings to 64-bit values with a MurmurHash
function.
\texttt{HashingFeaturizer} can be wrapped by other \Featurizer classes to
record vocabulary items and to exclude selected features from indexing.
By convention, features mapped to 0 are not indexed.
For example, the \texttt{JsonFeaturizer} wraps any \Featurizer,
and maps to 0 those tokens that represent JSON structural elements,
such as the curly braces surrounding objects.

An \Appender and an \Annotator work together for index construction and update.
Both support two-phase commit protocols,
with the overall transaction managed by the \Warren.
An \Appender appends data to the content through its \texttt{append} operation:
\[
\texttt{append(}\textit{content}\texttt{)} \rightarrow (p, q)
\]
The \texttt{append} operation returns the interval where the appended content
is located.
An \Annotator adds an annotation to the index through the
\texttt{annotate} operation:
\[
\texttt{annotate(}f, v, p, q\texttt{)}
\]
which adds the annotation $\left<f, (p, q), v\right>$,
where the value $v$ is optional.

\begin{figure}
\begin{minipage}[t]{0.45\textwidth}
{\small\tt
\begin{verbatim}

{
 "id": "0001",
 "type": "donut",
 "name": "Cake",
 "ppu": 0.55,
 "batters":
  {
   "batter":
    [
     { "id": "1001",
       "type": "Regular"},
     { "id": "1002",
       "type": "Chocolate"},
     { "id": "1003",
       "type": "Blueberry"},
     { "id": "1004",
       "type": "Devil's Food"}
    ]
  },
 "topping":
  [
   { "id": "5001",
     "type": "None"},
   { "id": "5002",
     "type": "Glazed"},
   { "id": "5005",
     "type": "Sugar"},
   { "id": "5007",
     "type": "Powdered Sugar"}
   { "id": "5006",
     "type": "Chocolate with Sprinkles" },
   { "id": "5003",
     "type": "Chocolate" },
   { "id": "5004",
     "type": "Maple" }
  ]
}
\end{verbatim}
}
\end{minipage}
\hspace*{-1cm}
{\small\tt
\fbox{
\begin{minipage}[t]{0.53\textwidth}
transaction()\\
append(\{) $\rightarrow$ (0, 0)\\
append("batters":) $\rightarrow$ (1, 4)\\
append(\{) $\rightarrow$ (5, 5)\\
append("batter":) $\rightarrow$ (6, 9)\\
append([) $\rightarrow$ (10, 10)\\
append(\{) $\rightarrow$ (11, 11)\\
append("id":) $\rightarrow$ (12, 15)\\
append("1001") $\rightarrow$ (16, 18)\\
annotate(:batters:batter:[0]:id:, 16, 18)\\
append(,) $\rightarrow$ (19, 19)\\
append("type":) $\rightarrow$ (20, 23)\\
append("Regular") $\rightarrow$ (24, 26)\\
append(\}) $\rightarrow$ (27, 27)\\
annotate(:batters:batter:[0]:type:, 24, 26)\\
annotate(:batters:batter:[0]:, 11, 27)\\
\textit{...}\\
annotate (:batters:batter:, 10, 84, 4)\\
\textit{...}\\
append("name":) $\rightarrow$ (95, 98)\\
append("Cake") $\rightarrow$ (99, 101)\\
annotate(:name:, 99, 101)\\
append(,) $\rightarrow$ (102, 102)\\
append("ppu":) $\rightarrow$ (103, 106)\\
append("0.5500") $\rightarrow$ (107, 110)\\
annotate(:ppu:, 107, 110, 0.55)\\
\textit{...}\\
annotate(:, 0, 254)\\
ready()\\
commit()
\end{minipage}
}
}
\caption{
Constructing an annotative index.
The inset on the right shows a partial trace of
\texttt{append} and \texttt{annotate} operations during the addition of the
JSON object on the left.
}
\label{fig:construction}
\end{figure}

Figure~\ref{fig:construction} shows a partial trace of \texttt{append} and
\texttt{annotate} operations,
while adding a nested JSON object to an annotative index.
With the help of a fast JSON parser\footnote{
\url{https://github.com/nlohmann/json}
},
support for a general JSON store requires less than 500 lines of C++ beyond the
core generic annotative indexing code.
The example object is taken from a set of open source examples available on
Adobe's website\footnote{
\url{https://opensource.adobe.com/Spry/samples/data_region/JSONDataSetSample.html}}.
The order that JSON key-value pairs are added differs from the textual
order in the object because the object is first parsed into a C++ map
and then traversed to add the object to the annotative store\footnote{
\url{https://github.com/claclark/Cottontail/blob/main/src/json.cc}}.

In the figure, 
the operation ``\texttt{append("batters":)}''$\!$ appends four tokens:
``\texttt{"}''$\!$, ``\texttt{batters}'', ``\texttt{"}'', and ``\texttt{:}'',
returning the interval $(1, 4)$.
Tokens marking structural elements of the JSON object
( ``\texttt{\{}'', ``\texttt{\}}'', ``\texttt{"}'', ``\texttt{:}'', etc.)
are encoded as special tokens using Unicode noncharacters,
which are permanently reserved for the internal use by systems that store and transmit text.
With this encoding,
the \texttt{translate} operation of a \Txt component,
which implements ${\cal X}(p, q)$,
can return any interval of the content and recognize the difference between a
``\texttt{:}'' separating a JSON key-value pair and a ``\texttt{:}'' that happens
to appear in a string.

For conciseness,
the trace omits \texttt{annotate} operations that add annotations for single
tokens, which are automatically performed as part of an
\texttt{append} operation.
For example, as part of the ``\texttt{append("Regular")}'' operation,
the annotation $\left<\texttt{regular}, 35\right>$ is automatically added.
As previously mentioned,
\texttt{JsonFeaturizer} returns 0 for tokens marking structural elements,
suppressing automatic annotation to avoid unnecessary indexing. 

All structure and nesting is retained in the features.
For example, the annotation
$\left<\texttt{:batters:batter:[0]:type:}, (24, 26)\right>$
indicates the ``\texttt{type}'' property of the first element of 
the ``\texttt{batter}'' array of the ``\texttt{batters}'' property.
A JSON object is not ``flattened'' in any sense.
The content (i.e, ${\cal X}(0, 254)$) contains the full JSON object,
which can be accessed by the \texttt{translate} operation of the \texttt{txt}
component.

By convention, the feature ``\texttt{:}'' is used as the root of the object,
as seen in the annotation $\left<\texttt{:}, (0, 254)\right>$.
Individual objects in a collection of JSON objects,
e.g. a JSONL file, can be accessed through this  `\texttt{:}'' feature.
In the annotation
$\left<\texttt{:batters:batter:}, (10, 84), 4\right>$
the value 4 gives the length of the array.
In a later example, we apply the convention of storing the array length as the
value for the array feature to step through (``explode'') arrays of different lengths in
different objects.
These conventions, as well as other conventions used to support a JSON store,
are independent of the underlying associative index.

\vspace{0.2cm}
\section{Query Processing}
\label{sec:processing}

The \Tau and \Rho access methods,
as defined by Equations~\ref{eqn:tau} and~\ref{eqn:rho},
provide the foundation for query processing.
In Cottontail, the \texttt{hopper(f)} operation of the \Idx component
creates a \texttt{Hopper} object for the 64-byte feature value \texttt{f}.
A \Hopper object acts as a cursor,
supporting the \Tau and \Rho access methods over the feature and caching
the most recent result from each access method.
All accesses to the underlying index structures
are abstracted by \Tau and \Rho,
which we are then free to implement in any suitable way.
For example, the index structures might include synchronization points to
allow the \texttt{Hopper} to skip annotations \citep{mz96}.
The current version of Cottontail represents annotation lists as arrays,
compressed until active,
and skips annotations with galloping search.
However, since the index structures are known only to the \Idx component,
it could employ any file structures and storage strategies able to
efficiently support the \Tau and \Rho access methods\footnote{
The name ``Cottontail'' was inspired by the ability of the
\Tau and \Rho access methods to efficiently ``hop'' around the index.}.

The translation function ${\cal X}(p, q)$ is implemented by the
\texttt{translate(p, q)} operation of the \Txt component.
A typical query processing loop for a structural query expressing containment
relationships might start with a query $Q$ expressed by the operators of
Figure~\ref{fig:operators}.
Calls to \Tau or \Rho generate successive solutions,
with the content translated and the results aggregated as needed.\\
\begin{minipage}[t]{4in}
  \begin{tabbing}
    \ \ \ \ \ \ \ \ \ \ \ \=\ \ \ \=\ \ \ \= \ \ \ \= \ \ \ \= \\ \kill
    \> {\tiny 1} \> $\mbox{\em Solve}(Q) \equiv$ \\
    \> {\tiny 2}  \> \> $\langle (p, q), v \rangle \leftarrow Q.\Tau(0)$ \\
    \> {\tiny 3} \> \> {\bf while}\  $p \neq \infty$:\\
    \> {\tiny 4} \> \> \> \mbox{\em Translate/Aggregate}\ $\langle (p, q), v \rangle$ \\
    \> {\tiny 5} \> \> \> $\langle (p, q), v \rangle \leftarrow Q.\Tau(p + 1)$ \\
  \end{tabbing}
\end{minipage}\\
The access methods return $\left<(\infty, \infty), 0\right>$ to indicate the end of the list.
As the solutions are generated,
the \Tau and \Rho operators allow solutions to subqueries to be
skipped when they cannot lead to a solution for the overall query.
The specific translation (${\cal X}$) and aggregation operations required on line~4 depend on the problem at hand.
Aggregations include the standard SQL aggregations (MAX, MIN, COUNT, etc.), which need to be preformed in memory.

\begin{figure}[htb!]
\centering 
\begin{tabular}{l|l|r|r}
    \textbf{Data Set} & \textbf{Description} & \textbf{Records} & \textbf{Size} \\ \hline
    \texttt{books} & Descriptions of technical books & 431 & 524K \\
    \texttt{city\_inspections} & Results of NYC business inspections & 81,047 & 23M \\
    \texttt{companies} & Overviews of tech companies & 18,801 & 74M \\
    \texttt{countries-big} & Country names by language & 21,640 & 2291K \\
    \texttt{covers} & Book ratings & 5,071 & 470K \\
    \texttt{grades} & Grades for homework assignments & 280  & 91K\\
    \texttt{products} & Phone and cable products & 11 & 2K \\
    \texttt{profiles} & Update log records  & 1,515 & 454K \\
    \texttt{restaurant} & Restaurant addresses and ratings & 2,548 & 666K \\
    \texttt{students} & Student grades & 200 & 34K \\
    \texttt{trades} & Stock trades & 1,000,001 & 231M \\
    \texttt{zips} & NYC zip codes & 29,353 & 3107K \\
    \hline
    \textbf{Total} & & \textbf{1,160,898} & \textbf{337M}
\end{tabular}
\caption{
Curated collection of heterogeneous JSON objects compiled by \"Ozler as a resource for exploring MongoDB
(\url{https://github.com/ozlerhakan/mongodb-json-files}).}
\label{fig:json-collection}
\vspace{0.5cm}
\end{figure}

To provide more concrete examples, we use the heterogeneous collection of
JSON objects presented in Figure~\ref{fig:json-collection}.
We base our examples on this collection due to its level of heterogeneity
and its independence from this work.
The collection was originally created as a resource for exploring and learning
MongoDB.
Compared to standard benchmarking tools \citep{brsr22} it provides a
reasonable source of clear and simple examples,
with an emphasis on heterogeneity.
Single-thread build time for this collection is just over 4 minutes for a
static index and just over 3 minutes for a dynamic index.

\noindent
Figure~\ref{fig:json-queries} presents these examples.
The Cottontail repo contains associated source code\footnote{
\url{https://github.com/claclark/Cottontail/blob/main/apps/json-examples.cc}
}.
For each query, we give a description in English,
a description in an SQL-like notation,
and a query in the structural query notation of Figure~\ref{fig:operators}.
The source code should be consulted for full details.
The figure includes query execution times for both static and dynamic indexes.

Examples 1-3 follow the general pattern above, i.e. a single query with
different types of aggregation.
Example 4 involves exploding an array containing author names.
In the figure, the structural query for this example returns each
array of author names as a whole,
while the example code in the repo illustrates the use of array indexes
to access individual elements one at a time.
Example 5 requires roughly a second on both indexes.
Processing this query requires over 80,000 accesses to the content,
corresponding to an average access time of $20 \, \mu s$ on the static index.
Even with various caching methods in place, there are limits on random
access to compressed text.
As much as possible, query processing should take place over the annotations.

The ``\texttt{FROM *}'' notation in Examples~7 is not valid SQL.
If it were, it would imply a Cartesian product of all tables.
Here, it suggests the ability to run queries that span objects with
different schema.
Examples~8 and~9 provide a more substantial example of annotative indexing
that enables unified queries over objects with different schema.
The objects in many of the subcollections include properties indicating
their creation date.
For example, in the \texttt{city\_inspections} subcollection, dates are
specified in a human readable format
(e.g, \verb|{"date":"Feb 20 2015"|).
In the \texttt{companies} subcollection,
some dates are specified as UNIX timestamps in milliseconds
(e.g. \verb|"created_at" : { "$date" : 1180075887000 }|).
With annotative indexing,
we can annotate the objects to provide consistent date annotations,
allowing Example~9 to count the objects created on a specific date
across all subcollections.

\begin{figure}
\vspace{0.5cm}
\centering
\begin{tabular}{l|r|r}
& \multicolumn{1}{c|}{\textbf{Static}} & \textbf{Dynamic} \\
\hline \hline
{\bf Example 1}: Statistics for restaurant ratings &  & \\
\hspace*{0.5cm}
{\small\tt
  SELECT MIN(rating), AVG(rating), MAX(rating)} & &\\ 
 \hspace*{1cm}
  {\small\tt FROM restaurant} & 14 ms & $<$1 ms \\
\hspace*{0.5cm}
$\mbox{\tt :rating:} \ContainedIn \mbox{\tt Files/restaurant.json}$ & & \\
\hline
{\bf Example 2}: How many zip codes does New York have? & &\\
\hspace*{0.5cm}
{\small\tt
  SELECT COUNT(*) FROM zips WHERE CITY = "NEW YORK"} & 23 ms & 2 ms\\
\hspace*{0.5cm}
$(\mbox{\tt :city:} \Containing
  \mbox{\tt "New York"}) \ContainedIn \mbox{\tt Files/zips.json}$ & & \\
\hline
{\bf Example 3}: Names of nanotech companies & &\\
\hspace*{0.5cm}
{\small\tt SELECT name FROM companies} & & \\
\hspace*{1cm}
  {\small\tt WHERE category\_code CONTAINS "nanotech"}
  & 133 ms & 3 ms \\
\hspace*{0.5cm}
$\mbox{\tt :name:} \ContainedIn
  (\mbox{\tt :} \Containing
    (\mbox{\tt nanotech}$ & &\\
    \hspace*{1cm}$\ContainedIn (\mbox{\tt :category\_code:}
      \ContainedIn \mbox{\tt Files/companies.json})))$ & & \\
\hline
{\bf Example 4}: Titles and authors of books & & \\
\hspace*{0.5cm}
{\small\tt
SELECT title, EXPLODE(authors) AS author FROM books} & 95 ms & 21 ms\\
\hspace*{0.5cm}
$(\mbox{\tt :title:} \OneOf
  \mbox{\tt :authors:}) \ContainedIn \mbox{\tt Files/books.json}$  & & \\
\hline
{\bf Example 5}: How many stock trades? & &\\
\hspace*{0.5cm}
{\small\tt
  SELECT COUNT(*) FROM trades} & 70 ms & 71 ms \\
\hspace*{0.5cm}
$\mbox{\tt :} \ContainedIn \mbox{\tt Files/trades.json}$ & & \\
\hline
{\bf Example 6}: Outcomes from city inspections & &\\
\hspace*{0.5cm}
{\small\tt SELECT result, COUNT(result) FROM city\_inspections} & & \\
{\small\tt \hspace*{1cm}GROUP BY result} & 1,686 ms & 939 ms\\
\hspace*{0.5cm}
$\mbox{\tt :result:} \ContainedIn \mbox{\tt Files/city\_inspections.json}$ & & \\
\hline
{\bf Example 7}: How many objects in the database? & & \\
\hspace*{0.5cm}
{\small\tt
  SELECT COUNT(*) FROM *} & $<$ 1 ms & $<$ 1 ms\\
\hspace*{0.5cm}
\texttt{:} & &\\
\hline
{\bf Example 8}: Titles of books publised in 2008 & &\\
\hspace*{0.5cm}
{\small\tt SELECT title FROM books} & &\\
{\small\tt\hspace*{1cm}WHERE created >= '2008-01-01'} & 13 ms & 9 ms\\
{\small\tt\hspace*{1cm}AND created <= '2008-12-31'} & &\\
\hspace*{0.5cm}
$\mbox{\tt :title:}
  \ContainedIn (\mbox{\tt Files/books.json} \Containing \mbox{\tt year=2008})$ & & \\
\hline
{\bf Example 9}: Count objects created on December 1, 2008. & & \\
\hspace*{0.5cm}
{\small\tt SELECT COUNT(*) FROM * WHERE created = '2008-12-01'} & 13 ms & 4 ms \\
\hspace*{0.5cm}
$\mbox{\tt :} \Containing
  (\mbox{\tt year=2008} \BothOf \mbox{\tt month=12} \BothOf \mbox{\tt Day=01})$
  & & \\
\end{tabular}
\caption{
Illustrative examples of containment and other operations over the JSON collection
from Figure~\ref{fig:json-collection},
with query processing times over static and dynamic index structures.
Examples~8 and~9 depend on additional date annotations not present in the
original JSON.
The SQL queries are provided for explanatory purpose;
they cannot be directly executed by the reference implementation.
The structural queries describe index access only;
additional processing is required to complete query processing,
including aggregations.
}
\label{fig:json-queries}
\end{figure}

\vspace{0.2cm}
\section{Dynamic Update}
\label{sec:update}

Annotative indexing fosters a dynamic view of the content it stores.
After we append text to the content,
we can annotate it in different ways and for different purposes.
For example, the transformations applied to the MS MARCO corpus described in
the introduction, including tagging and segmentation into passages,
could be achieved through annotations. 
For ranking purposes, term frequency values at the document level
can be combined with sparse learned weights at the passage level to support
hybrid search.
Fields in heterogeneous collections of objects can be unified and
related objects can be linked.

This section outlines an approach to dynamic update of an annotative index
that maximizes flexibility,
including support for multiple simultaneous readers and writers.
Updates are grouped into transactions.
At the start of a transaction, a snapshot is taken of the index state,
which remains active until the transaction is committed or aborted.
Both content and annotations in this snapshot can be accessed on read-only
basis until the transaction ends.
For example, during the transaction we might read the content to identify
sentence boundaries in passages, or to compute term statistics.
During the transaction, we can append to the content and add annotations,
but these changes will not be immediately visible in the snapshot.
We can also erase content and annotations.
Once the update is complete,
we follow a two-phrase protocol to commit or abort the update,
allowing us to support transactions that span independent annotative indices.
After the transaction is complete,
the updated content and annotations become visible.

While the Cottontail's static index supports only one transaction at a time,
its dynamic index supports multiple concurrent transactions.
Each transaction is managed by a \Warren (see Figure~\ref{fig:cottontail}).
The \texttt{clone} operation allows a \Warren to be copied for the purpose
of supporting concurrent transactions,
with each clone managing one transaction at a time.
For example, in a multi-threaded application each thread could \texttt{clone}
a copy for it own use.
The \texttt{start} operation captures the read-only snapshot of the index,
while the \texttt{end} operation releases this snapshot.
Any accesses to the \Warren, even read-only access,
must be bracketed by a \texttt{start}/\texttt{end} pair.
The \texttt{transaction} operation starts a write transaction,
at which point the \Appender and \Annotator may be used.
In addition, to the \texttt{annotate} operator, 
the \Annotator supports an \texttt{erase} operation that removes the content
and its annotations over a specified interval by annotating the interval
with the reserved feature 0.
\Txt and \Hopper objects skip these intervals until the associated content
and annotations can be garbage collected.
The remaining operations~---~\texttt{ready}, \texttt{commit},
and \texttt{abort}~---~complete the two-phase commit protocol.
The update is not visible to the \Warren until after the \texttt{end} operation,
followed by another \texttt{start}.

Internally, each committed transaction creates a special update \Warren
object that contains only the newly added content and annotations\footnote{
In a dynamic index, warrens multiply like rabbits.
}.
After a commit, an update \Warren object is immutable.
At the start of the ready phase of the two-phase commit,
the index assigns an update \Warren a sequence number.
A vector of \Warren objects in sequence order provides the snapshot used
for read access.
In the background \Warren objects are merged and garbage collected,
with a merged \Warren representing a subindex of the full index,
corresponding to a range of updates in sequence order.
Once a \Warren is merged into a larger range and is released from all active
snapshots, it is deleted.

\noindent
During an update,
content and annotations are assembled in a separate address space.
At the start of the ready phase,
when the index knows the final length of the appended content,
it assigns a permanent address interval to the content and maps newly
added annotations to this interval.
During the ready phase the update is also logged durably to storage.
If the commit is aborted after the ready phase,
the assigned address interval becomes a gap,
and the update is garbage collected from the log.
During the update process, a global lock is held only for brief periods,
such as when a snapshot is taken or when sequence numbers and address
intervals are assigned.

Cottontail supports ACID properties of transactions.
Transactions are fully atomic,
with newly added content and annotations remaining invisible to \Txt and \Idx
operations until the transaction is committed.
Cottontail guarantees consistency in that updates to annotations preserve
minimal interval semantics.
However, to maximize concurrency,
Cottontail provides limited support for isolation.
If concurrent transactions add annotations for the same feature that nest,
the index retains only the innermost.
If concurrent transactions add annotations with the same start and end
addresses, the index retains only the value from the one with the largest
sequence number.
A failure before the start of a final commit phase of a two-phase commit,
guarantees that the transaction is aborted, with no changes.
A failure after a commit guarantees that the update is durably recorded.
A failure during commit processing will leave the index in a consistent state,
with the transaction either committed or aborted.

\begin{figure}[htb!]
\centering
\includegraphics[width=\textwidth, keepaspectratio]{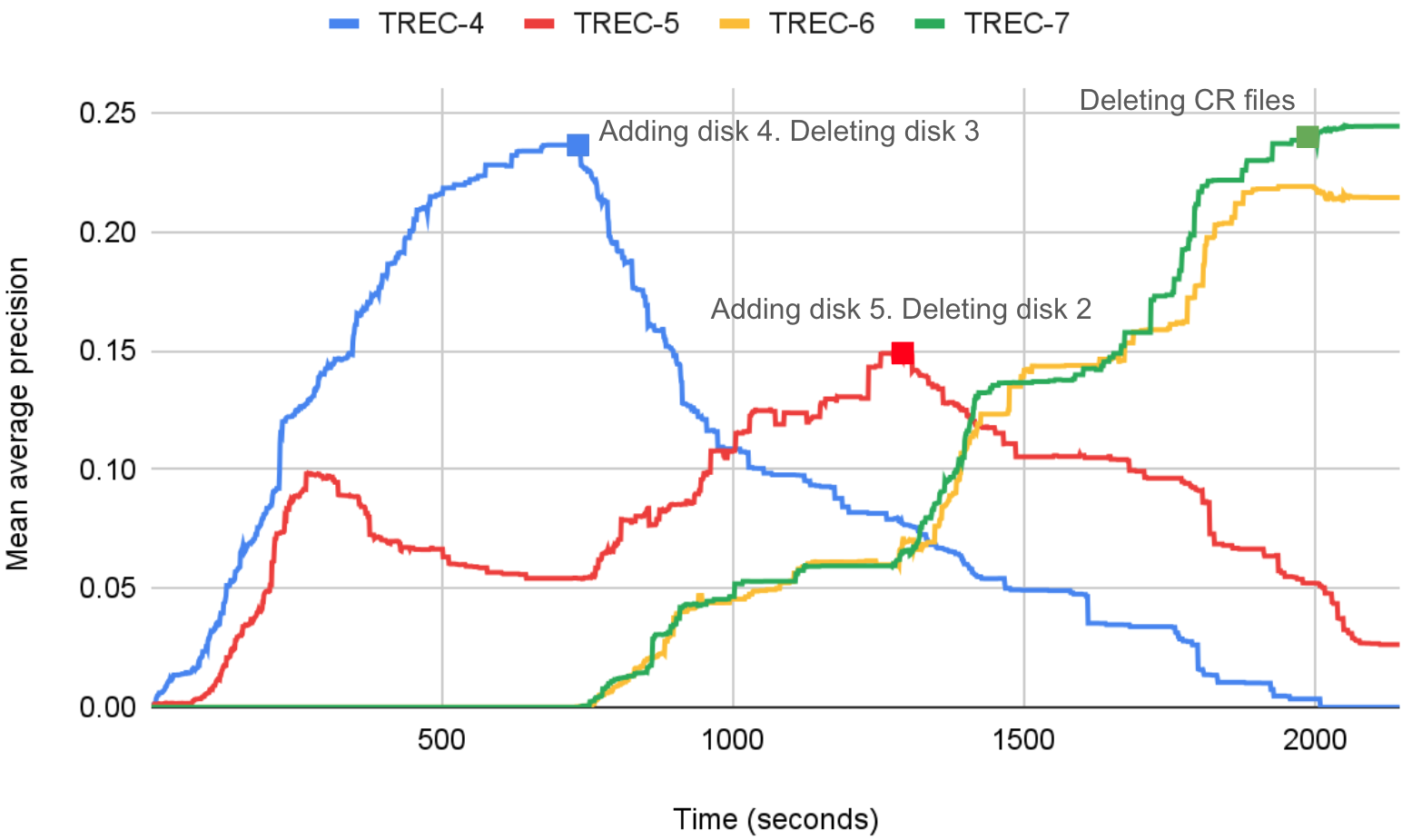}
\caption{
Example of transaction processing in cottontail.
The example recapitulates four years of early TREC experiments,
when the collection was changing significantly from year to year.
The example was generated by 28~appending
threads~---~one for each processor core~---~one deletion thread,
and 199~querying threads~---~one
for each query in the TREC-4 to TREC-7 test collections.
The appending threads append each of the 4,905 files in the
TREC collection as a separate transaction.
They then add ranking statistics and relevance information as
separate transactions.
The deletion thread removes documents, so that collection evolves from year to year. 
The querying threads run continuously,
each executing a BM25 query with pseudo-relevance feedback and then computing
mean average precision using relevance information from the index.
The lines in the figure plot MAP values as they change over the course of the experiment.
}
\label{fig:transactions}
\vspace{0.5cm}
\end{figure}

Figure~\ref{fig:transactions} proves an illustration of dynamic update with
multiple concurrent readers and writers\footnote{
\url{https://github.com/claclark/Cottontail/blob/main/apps/trec-example.cc}
}.
The figure recapitulates four years of older TREC experiments when the
test collection changed substantially from year to year \citep{trec7}.
Documents for the test collection were distributed on five disks,
encoded in an HTML-like format and organized into 4,905 files.
TREC-4 used disks~2 and~3; TREC-5 used~2 and~4; TREC-6 used~4 and~5;
TREC-7 dropped the low quality CR subcollection from disk~4.
50~new queries were introduced each year,
but one query was excluded from TREC-4, leaving 199 queries in total.
Each query was judged for relevance over the collection from the year it was
introduced, with an average of 1,866 judgments/query.
The figure was generated by hundreds of threads concurrently reading and
writing a Cottontail dynamic index, including:
\begin{enumerate}
\item
  28 appending threads, one for each core.
  Together they append the entire collection, a file at a time.
  Each file is appended as a separate transaction.
  After each append is committed,
  the thread re-reads the documents from the index,
  computes term statistics for them,
  and writes the statistics to the index as a second transaction.
  Finally,
  if there are documents in the file that are relevant to any of the queries,
  annotations reflecting these relevance judgments are written to the index
  as annotations in a third transaction.
\item
  199 querying threads, one for each query.
  Each repeatedly starts a read access,
  runs its query with BM25,
  expands the query using pseudo-relevance feedback over the top~20 documents,
  runs the expanded query to return the top~1000 documents,
  reads relevance judgments from the index,
  computes average precision,
  and reports it on output where it is captured for later summarization on a per-year basis.
\item
  One deletion thread.
  It erases documents, a file at a time,
  so that the collection evolves over time.
  Each file is erased as as separate transaction.
  The squares in the figure indicate points where the deletion thread
  synchronizes with the other threads so that all queries are executed at least
  once on the entire collection for a given year.
\end{enumerate}
In addition to these application-level threads, 
maintenance threads work throughout the experiment to merge and garbage
collect the index.
The experiment requires 16,442 update transactions in total.
By the end of the experiment,
these have been merged into 12 subindexes,
each corresponding to a thousand or so sequence numbers.
Throughout the experiment,
processor utilization essentially remains at 100\% on all cores.

As documents are added to the index for a given year,
the MAP value for that year increases until it hits a synchronization point.
It then drops as documents are deleted.
The BM25 parameters are tuned for more recent collections.
The peak MAP values represent good performance on TREC-6 and TREC-7,
and reasonable performance on TREC-4 and TREC-5.

\section{Conclusion}
\label{sec:ongoing}

This paper introduces and explores annotative indexing,
a novel and flexible indexing framework, which unifies and generalizes inverted indexes, column stores, object stores, and graph databases.
A particular feature of annotative indexing is its ability to manage heterogeneous collections of semi-structured data, unifying common elements across diverse formats.
Text in any format can simply be appended to the content,
with annotations added at a later time for a variety of purposes,
such as sentence segmentation, tagging, or indexing for ranked retrieval.

Integrating annotative indexing into a retrieval augmented generation (RAG) system \citep{rag} forms a primary focus for current and future work.
Given a few examples, a large language model (LLM) can generate structural queries using the operators of Figure~\ref{fig:operators}, allowing natural language queries to be translated into structured queries over heterogeneous content.
For example, imagine a life-logging application supported by a RAG system that integrates an annotative index.
Messages, mail, conversations, and other experience could be poured into the index as content for ongoing tagging, linking, indexing, and other annotation.
From the perspective of a person using the application, querying their past experience (``I really liked the movie I saw on the plane last weekend. What are similar movies I haven't seen yet?'') happens in natural language, but internally this query could be handled by a combination of ranked retrieval and structured queries to a knowledge graph linked with the experiences. 

Extending annotative indexing to support dense retrieval provides another immediate goal.
While a 64-bit value in an annotation cannot store a dense vector,
it can store a vector identifier.
However, to better support a fully dynamic index,
the author plans to mimic the approach taken for the content translation function ${\cal X}(p, q)$ by associating vectors with positions in the address space.
A vector mapping function ${\cal V}(p)$ would return the vector associated with a location in the address space, presumably the location where the corresponding content appears. 
In this way, dense vectors can be garbage collected as intervals in the address space are erased.

To provide further support for dense retrieval, current work also includes an exploration of methods for encoding HNSW graphs as annotations.
Graph structures can be represented in two ways by an associative index.
First,
as suggested in Section~\ref{sec:links},
we can store an address as the value in an annotation,
so that $\left<G, p, v\right>$ indicates a directed edge from $p$ to $v$ 
in the graph $G$.
However, unless we are careful with updates,
this representation can create ``dangling references'' to deleted content.
An alternative representation stores a feature representing a list of
out edges as the value in an annotation.
Under this representation,
the value in the annotation $\left< G, p, E_p\right>$ is a feature indicating outlinks from the content at $p$ in the graph $G$.
An annotation for $E$ of the form $\left< E_p, p'\right>$ indicates a directed edge from $p$ to $p'$.
While the details are left for a future paper,
this second representation should allow for the representation and traversal of HNSW graphs through annotations.

\noindent
At the time of writing, the largest collection indexed by Cottontail is the 350GB C4 corpus,\footnote{\url{https://huggingface.co/datasets/allenai/c4}} which the author routinely uses for cross-collection pseudo-relevance feedback.
Ongoing work includes scaling Cottontail to handle larger collections, as well as fully distributed collections.
Cottontail has indexed Wikidata as eight shards directly from its JSON dump,\footnote{\url{https://www.wikidata.org/wiki/Wikidata:Database_download}} and the author is exploring support for knowledge graph queries over this collection.

At the time of writing, Cottontail remains an experimental system.
Compilation requires the Bazel build system and the Boost C++ library, both of which require some effort to install.
The system has not been ported to Mac or Windows, and it currently runs only on Ubuntu.
In the near future, I plan a single-file release in the spirit of SQLite, along with examples illustrating common use cases.
Wider use of Cottontail also requires a Python wrapper.

Current performance on basic BM25 ranking does not quite match that of Lucene.
In the immediate future, I plan to focus attention on improving performance of the lowest level code for the \texttt{Hopper} operations, which should improve general performance, including ranking.
Finally, the only delete operation supported by Cottontail is to erase all content and annotations from an interval of the address space.
Additional delete operations might delete specific annotations or all annotations for a given feature.

\acks{%
The reference implementation for annotative indexing has its roots as a pandemic project.
While I do not exactly thank the pandemic, I appreciate it gave me time to do some things that I would not otherwise have time to do.
I made a final push to complete this work while was on a six-month sabbatical May to October 2024.
During May and June, Yiqun Liu, Min Zhang, and Qingyao Ai kindly hosted me for a visit to Tsinghua University in Beijing.
During September and October, Craig Macdonald and Iadh Ounis kindly hosted me for a visit to the University of Glasgow.
Mark Smucker and Negar Arabzadeh read earlier versions of this paper and provided helpful feedback.
}

\vspace{0.2cm}

\bibliography{paper}

\end{document}